\documentclass[letterpaper,twocolumn,10pt]{article}
\usepackage{usenix2019_v3}

\usepackage{tikz}
\usepackage{amsmath}

\usepackage{filecontents}

  \usepackage{cite}
\usepackage{amsmath,amssymb,amsfonts}
\usepackage{algorithmic}
\usepackage{graphicx}
\usepackage{textcomp}
\usepackage{xcolor}

\usepackage{listings}
\usepackage{fancyvrb}
\usepackage{subfigure}
\usepackage[linesnumbered,boxed,ruled,commentsnumbered]{algorithm2e}
\usepackage{multirow}
\usepackage{hyperref}
\usepackage{appendix}
\usepackage{amsmath,amssymb,amsfonts}

\begin{document}
%

\title{\Large \bf SRAS: Self-governed Remote Attestation Scheme for Multi-party Collaboration}



\author{Linan Tian\\
tianlinan@intel.com
\and
Yunke She\\
shenyunke@intel.com
\and
Zhiqiang Li\\
richard.z.li@intel.com
}



%


\maketitle{}

\begin{abstract}
Trusted Execution Environments (TEEs), such as Intel Software Guard Extensions (SGX), ensure the confidentiality and integrity of user applications when using cloud computing resources.
However, in the multi-party cloud computing scenario, how to select a Relying Party to verify the TEE of each party and avoid leaking sensitive data to each other remains an open question.
In this paper, we propose SRAS, an open self-governed remote attestation scheme with attestation and verification functions for verifying the trustworthiness of TEEs and computing assets, achieving decentralized unified trusted attestation and verification platform for multi-party cloud users.
In SRAS, we design a Relying Party enclave, which can form a virtual verifiable network, capable of local verification on behalf of other participants' relying parties without leaking sensitive data to others.
We provide an open source prototype implementation of SRAS to facilitate the adoption of this technology by cloud users or developers.
\end{abstract}


%


\section{Introduction}
Cloud computing has become popular with its naturally cheap computation and storage strengths.
A vast number of applications are deployed in third-party data centers and public cloud environments, 
such as Alibaba Cloud \cite{Alibaba}, Google Cloud \cite{Google}, and Microsoft Azure \cite{Microsoft}.
However, cloud users and security researchers argue that there are potentially privacy issues with cloud providers' services 
and whether their offerings can guarantee the confidentiality of data from running applications and whether these applications execute as expected.
Trusted Execution Environments (TEEs) offer effective solutions to these issues, such as Intel Software Guard Extensions (SGX) \cite{SGX2013}.
As a result, significant efforts have been made over the past few years for broader TEE adoption
(e.g., \cite{hoekstra2013using,ohrimenko2016oblivious, schuster2015vc3, tamrakar2017circle, seo2017sgx, lind2017glamdring, baumann2015shielding, tsai2017graphene, priebe2018enclavedb, zhao2022vsgx, shen2020occlum, zhang2021succinct, chen2022sgxlock}).

However, in cloud computing multi-party scenarios, the above approaches are not effective in guaranteeing the trustworthiness of attestation and verification for TEE and compute asset.
As shown in Figure \ref{fig:per-party}, 
cloud users will challenge whether the privacy enclave is located in the TEE, i.e., to verify that it is running correctly on their own TEE platform or another party's one.
Subsequent operations can only continue once the attestation results have been accepted by all parties involved.
Prior to attestation, all privacy enclaves are required to provide or preconfigure sensitive information about each party for the relying party and verifier.
Including Independent Software Vendor (ISV) identity information (the participant to which the ISV belongs or its IP address), enclave identity (MRENCLAVE), ISV signer information (MRSIGNER) and Trusted Computing Base (TCB) information.
So that replying party can accurately verify that all involved parties need to run privacy enclaves.
As the result, multi-party cloud users have the following security concerns:

\begin{figure}
	\centering
	\includegraphics[width=0.48\textwidth]{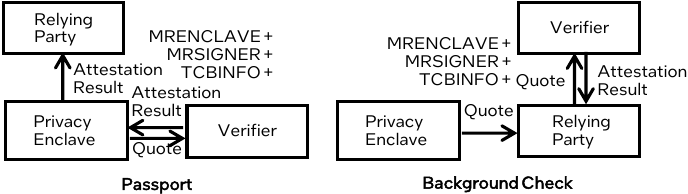}
	\caption{In passport pattern, Privacy Enclave sends the Quote to the Verifier for attestation and Relying Party verifies the attestation result. In background check pattern, Privacy Enclave first sends the Quote to Relying Party and Relying Party forwards the Quote to the Verifier, Relying Party then verifies the attestation result from the Verifier.}
	\label{fig:per-party}
\end{figure}

\noindent \textbf{How to vote for a relying party that each participant accepts to verify all TEEs.} 
For the single party, they can specify or design self-accepting relying party for remote attestation.
With the help of the third-party SGX Data Center Attestation Primitives (DCAP) \cite{dcap}, 
besides verifying the TEE of the privacy enclave, 
they verify the privacy enclave by setting up the identity of the privacy enclave in the relying party verification logic.
However, it is difficult to vote for a relying party accepted by all parties to verify all privacy enclaves including the TEE.
The credibility of the chosen relying party is difficult to measure because it should not be affiliated with any party.
In the case where the relying party is a participant (or the third-party), 
a co-location attack may be implemented once the relying party obtains the identity of the TEE platform and privacy enclave \cite{ristenpart2009hey, varadarajan2015placement}.

\noindent \textbf{How to guarantee that each participant's sensitive data will not be leaked to the others.} 
When verifying the identity information of the privacy enclave, the replying party needs to know privacy enclave's identity in advance.
Therefore, for multi-party cloud users, all parties' privacy enclaves identity information is gathered into the relying party.
Once the relying party exists vulnerable or untrustworthy (co-location attack), all parties' privacy information will be leaked. 
In addition, for that participant with special Intellectual Property which its TCB platform information also does not want to disclose to anyone.
Using a common relying party is even more dangerous.
Because all parties' TCB platform information is also known to the relying party.

Recognizing these security concerns, MAGE \cite{chen2022mage} has designed an attestation framework which enables a group of enclaves for mutual attestation without third-party relying party.
It proposes to split the enclave into two parts, 
one part retaining the original functional logic code and a common part deriving the identity of the same group of enclaves.
While this work partially mitigates the problem, its approach does not support enclave updates.
In order to store the identity information of the other parties, the common parts of the enclaves within the framework need to be developed and released centrally.
Among multi-party users, once a privacy enclave is updated, the other enclaves in that collaborative group will need to be redeveloped as well.
In addition, centralized development and distribution implies the need to find a relying party in the collaborative group.
This seriously deviates from security design principles (the security concerns described above).
Additionally, 
MAGE cannot anonymously verify that one party has launched a privacy enclave in an unexpected TEE.
Other work OPERA \cite{chen2019opera} proposes an open platform for remote attestation in single party enclave.
While this open platform guarantees that private local information is not leaked to third parties, 
it faces the same problem as MAGE in a multi-party context.

\begin{figure}
	\centering
	\includegraphics[width=0.475\textwidth]{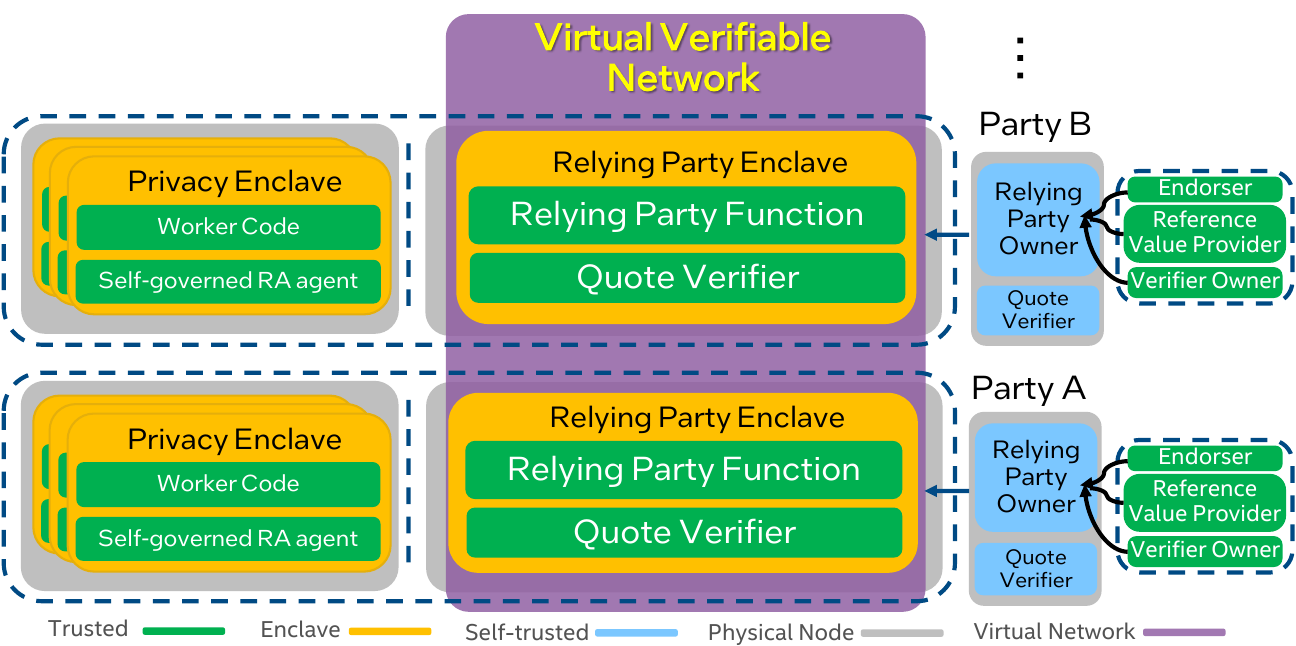}
	\caption{SRAS multi-party architecture overview. 
 Each participant is an RP owner, and each of them acts as a verifier owner to provide endorsements and reference values to build a consensus $policy$. 
 After attesting the local RPE, each party's trusted anchor is conducted by the RP owner to local RPE.
 The RPEs granted by each party establishe a virtual verifiable network through mutual attestation, 
 implementing the validation of the trustworthiness of multi-party attestation and verification.}
	\label{fig:overall}
\end{figure}

In this paper, we want to explore a different design for a lightweight remote attestation solution to guarantee the trustworthiness of multi-party attestation and verification in cloud computing.
We present the {\itshape Self-governed Remote Attestation Scheme (SRAS)}, an open-source attestation solution, which does not break the any structures of enclave.
The inspiration for SRAS is to build a Relying Party (RP), local to each participant, 
with each RP performing attestation locally on behalf of the other participant's RPs.
In other words, one participant's enclave verifying itself to its own RP is equivalent to verifying itself to the RPs of other parties.
All sensitive information in each party's privacy enclave does never leave locally.
In this way, guaranteeing the trustworthiness of the attestation.

As such, we design a RP Enclave (called RPE) that located in each party, which contains the relying party and verifier functions, 
as illustrated in Figure \ref{fig:overall}.
It conducts mutual attestation to form a virtual verifiable network.
In this network, RPE attests that the attestation results of the local privacy enclave can be verified by other parties' RPEs.
We extend DCAP Quote to embed public key generated by local RPE as RPE identifier of participant into the $report\_data$ of the Quote.
Since the MRENCLAVE is the same for all RPEs, 
after mutually attesting the RPE Quotes, 
each RPE can use the other parties' key identifier to verify the local privacy enclave attestation results of other parties.
In other words, each RPE located on each party is able to locally attest on behalf of the other parties' RPEs.
In this way, cloud users are not required to elect a relying party to do remote attestation in multi-party scenarios.

In order for each party to accept the attestation results of other parties, we design a common $policy$ strategy in SRAS.
This appraisal $policy$ is a consensus formed by all parties involved. 
We propose that each participant is a RP owner and can provide endorsements and reference values, but such information must be approved by all parties (documenting includes all collaborative platform TCBs and identity information, etc.).
We extend DCAP Quote to embed the hash of $policy$ during RPE mutual attestation to verify that the $policy$ strategy configured in each party's RPE is pre-negotiated.
We design that RPE to strictly follow the $policy$ strategy to perform all verification behaviors.
For example, based on the RPE TCB information recorded in the $policy$ strategy, 
each party's RPE verifies the correctness of Quote received from other parties.
Since the implementation of RPE is completely open, no one is able to launch an unknown enclave or modify the reference values used for attestation by violating the $policy$ strategy without prior approval.
In this way, each RPE located on each party is able to accept the local verification results from other parties.

We have implemented a prototype of SRAS using Intel SGX DCAP and third-party open-source Gramine.
Our evaluation suggests that SRAS can effectively prevent attackers from attacking Relying Party with a total authentication latency of less than 1 second.
The building time of the multi-party secure channel is only 87.6ms.

In summary, our main contributions are as follows:
\begin{itemize}
\item We present SRAS, a lightweight self-governed remote attestation scheme, which provides relying parties with attestation and verification functions to validate the trustworthiness of trusted execution environment and compute assets, achieving a decentralized unified trusted attestation and verification platform for multi-party cloud users. 
\item We design an open-source relying party enclave in SRAS, which forms a virtual relying party verifiable network that conducts the local trust of the participants into the RPE group network and is able to locally verifying on behalf of the relying parties of the other participants without leaking the sensitive data to others, guaranteeing the privacy and security of multi-party cloud computing.
\item We have implemented this scheme in our prototype, and systematically characterized its performance overhead. 
The results show that it has reasonable overhead for TEE attestation and verification and can be used in practice.
We have already open-sourced SRAS on https://github.com/M-Party/Self-governed-Remote-Attestation-Scheme.git.
\end{itemize}

We proceed as follows:
We provide the background on SGX and DCAP in Sec. \ref{background}.
We present the preliminary for this challenge and overview of proposed scheme in Sec. \ref{Pre_Over}. 
We describe the design details for SRAS in Sec. \ref{design}. 
Sec. \ref{implementation} presents the prototype implementation and Sec. \ref{evaluation} conducts security analysis and evaluation.
Sec. \ref{discussion}, Sec. \ref{relatedwork} and Sec. \ref{conclusion} discuss the application scenario of SRAS, related work and conclusion, respectively.

\section{Background}\label{background}
In this section, we present the background on SGX and DCAP that are necessary to further the understanding of our approach.

\subsection{SGX} 

SGX, a hardware-based trusted computing technology developed by Intel, 
serves the critical purpose of safeguarding sensitive data and code from unauthorized access and alterations, 
even in the presence of malicious software or hardware \cite{206170, costan2016intel}. 
This cutting-edge technology introduces a concept known as enclaves, 
specialized memory regions that are created and secured by the processor. 
Enclaves can only be accessed by trusted code running within them. 
Leveraging hardware features like Enclave Page Cache (EPC)\cite{fei2021security}, 
SGX effectively safeguards the code and data within enclaves against unauthorized access and tampering, 
even in cases where the operating system or other software components are compromised or malicious.

\subsection{DCAP}\label{bkg:DCAP} 



In order to ascertain the identity and integrity of the enclave, a practice known as remote attestation is of vital importance. 
In Intel SGX, remote attestation signifies the procedure for demonstrating that software executes on an Intel SGX-enabled platform securely ensconced within a properly initialized enclave. 
Attestation occurs before granting the software access to secrets and protected data. 
The latest remote authentication scheme supported by Intel SGX is the Data Center Attestation Primitives (DCAP). 
Within DCAP\cite{dcap}, the Provisioning Certification Service (PCS) and Quoting Enclave (QE) furnish a certificate chain rooted in an Intel-issued certificate. 
This chain is used to certify enclaves to facilitate remote attestation in non-Intel attestation infrastructures. 

To certify the platform ensconced within an enclave, a Report generated by the enclave will be transformed into a “Quote”. 
The quote encompasses vital information such as the enclave's measurement value, attributes, public key, and more, accompanied by a signature value signed by the Attestation Key within the QE. 
An Intel-provided enclave known as the Provisioning Certification Enclave (PCE) functions as a local Certificate Authority for local QEs, utilizing its private key to sign the Attestation Keys generated by QEs. 
Upon receiving the attestation public key of QE, the PCE authenticates it and issues a certificate identifying the QE and the Attestation Key. 
This certificate is signed by the Provisioning Certification Key (PCK). 
Certificates and certificate revocation lists (CRLs) for the PCKs are made accessible on all genuine Intel platforms, forming a complete signature chain from the Quotes to an Intel Certificate Authority.

Consequently, anyone possessing the full certificate chain and CRLs can verify the Quotes. 
PCK certificates and other platform collaterals, including CRLs and other platform TCB information, are loaded into the validator's Data Center Caching Service (PCCS) and are available for use in runtime attestation requests. 
To validate an Intel SGX attestation, the relying party should adhere to the following steps:
\begin{itemize}
\item Verify the integrity of the signature chain from the Quote to the Intel-issued PCK certificate.
\item Check for any revoked keys within the chain.
\item Confirm that the QE originates from a trustworthy source and is up to date.
\item Verify that the status of the Intel SGX TCB is described within the chain.
\item Ensure that the enclave measurements in the Quote align with the expected enclave identity.
\end{itemize}

In the process of DCAP remote attestation, having the necessary platform collaterals is crucial. 
The PCCS plays a vital role in obtaining this information from the PCS based on the platform ID. 
It is essential that the platform is pre-registered in the PCS to facilitate this retrieval process.

The verification steps are executed within a special enclave, signed by Intel, known as the Quote Verification Enclave (QVE). 
This design ensures that the entire process and the results of remote attestation are safeguarded against tampering or unauthorized access. 
Furthermore, the identity of the QVE can also be verified by the relying party, enhancing the overall security and trustworthiness of the remote attestation process.

\section{Preliminary and Overview}\label{Pre_Over}
In this section, we describe the threat model we assume.
We then list the challenges for solving multi-party attestation and verification problem, and provide key insights on how SRAS addresses these challenges.
We also outline the workflow of our remote attestation scheme.

\subsection{Threat Model}\label{threat}
The goal of our work is to guarantee the trustworthiness of multi-party attestation and verification in cloud computing.
The only trusted anchor for each participant is local root secrets in itself intranet, and other parties are untrusted.
But, the security of entire software stack is out of scope.
We assume all unattested and unauthorized SGX-enabled enclaves are untrusted but are secure.
The rest of software stack, including the host applications, OS and hypervisor is consider as untrusted.
They might suffer from unprivileged, authorized or system software adversary attacks.
Additionally, we assume the adversaries are also able to control the all communication network.
They can perform man-in-the-middle, intercepting, dropping, replaying, and other attacks against the communication protocols between enclaves and software.

We assume that malicious users and software shall be prevented from getting access to data and codes inside the enclaves.
Therefore, the hardware implementation of TEE is secure, an adversary cannot breach the confidentially and integrity of enclave code and data.
We assume the third enclaves provided by Intel are logically sound and secure, and the TCB platform information is trust.
The memory corruption attacks \cite{biondo2018guard, lee2017hacking, van2019tale} occured in enclave or side-channel attacks \cite{chen2019sgxpectre, van2019ridl, brasser2017software, gotzfried2017cache, hahnel2017high, lee2017inferring, schwarz2017malware, van2017telling, wang2017leaky} are out of the scope.
But the adversaries are able to launch an enclave as they want.
We do not consider denial-of-service (DoS) attacks. 
Defending against such attacks are orthogonal to our goals.

\subsection{Challenges and Key Insights}\label{Insights}

\noindent \textbf{C1: How to establish a reliable relying party for the verification.} 
Before the multi-party take collaboration based on TEE related cloud computing, 
participant needs to verify whether the counterpart's TEE is legal and whether the privacy enclave has been launched correctly.
For the centralized attestation schemes, all participants trust a third relying party, and which attests all participants' privacy enclaves.
The privacy enclaves could communicate with the third relying party to obtain attestation results of collaborated enclaves. 
For all parties, the third relying party might suffer a single point of failure,
leading to unavailability of the TEE attestation service. 
Additionally, the verification service operators still face liability issues, 
which might engage in co-location attacks in conjunction with a malicious participant \cite{varadarajan2015placement}. 
As a result, third relying parties also face the problem of meeting regional and national regulatory requirements.

\textbf{SRAS enables a decentralized relying party without trusted third-party.} 
We can strategically avoid the need a third relying party if we only do local attestation for each participant's enclaves so that each participant's trust does not leave the local party.
Since SRAS's attestation service is completely open and publicly verified, 
we are able to effectively avoid the local relying party co-work with Cloud Computing Provider to implement a coordinated deception.
For each participant, the root of trust is only itself, so we design a Relying Party Enclave, RPE to inherit the root of trust of each participant.
Each participant first attests the RPE's TEE correctness, then delegates RPE to verifies all privacy enclaves in cloud computing devices.
This allows us to avoid to establish a third relying party for attestation. 

\noindent \textbf{C2: How each party's verification results are recognized by others.} 
For the centralized attestation schemes, since each participant trusts the third relying party, 
they could query the verification results from the third relying party.
If the results notified by third relying party are valid, one party recognizes that the others have been successfully verified.
However, as discussed above, how to establish a reliable relying party is a challenge in itself.
For the decentralized attestation schemes, it is a natural gap to make other parties recognize the results of their own verification.
If each participant delegates trusted local relying party to verify all of others, 
it needs to provide others' attested identities and TCB platform information in advance, which will leak participant's sensitive information.
Some work \cite{chen2022mage} attempt to modify the enclave structure to store others' identify to implement automaully mutually attestation.
However, this approach also needs a third party to unified develop and release all participating enclaves.

\textbf{SRAS abstracts the relying party to form a virtual mutual verification network, implementing results verification.} 
Unlike other efforts, we observe that {\itshape for each local relying party, if it is able to delegate other parties' relying party, which attests is equivalent to other parties do attestation,
the local verification results also are accepted by others.}
Therefore, if RPEs are able to record some endorsers by mutual attestation between all relying party enclaves, using recorded endorsers to verify the verified results of each party, the local verification results could be recognized.
Since all RPEs code and data in the enclave section are the same, we do not record identity for all parties' RPE to mutual attestation in advance.
After careful examination of DCAP Quote structure \cite{dcap}, we find only extent the $report\_data$ of the Quote to embed public key certificate generated by local RPE, can be used as an endorsement for other RPEs' results verification.
To this end, before each RPE attests the local privacy enclaves, all RPEs form a virtual verifiable network to pass the Quote to take the mutual attestation.
If all RPE Quotes verification success each other, each party's RPE records the other party's public key certificates.
When the local privacy enclave verification results update to other RPEs, each party can use record public key certificates to verify these results signed by other RPEs.

\noindent \textbf{C3: How to guarantee that all parties' TEEs are properly configured.}  
There is a coordinated deception between the relying party and some participants.
If TEE of launched privacy enclave in one party exists vulnerable, 
which can combine the third relying party to deceive other party to obtain privacy data in its privacy enclave memory.
Additionally, even though the third relying party is a trusted party, existing a system software adversary in the relying party, 
which might modify the stored TCB information during the validation process to pass the malicious TEE' enclave Quote, 
such as modifying PCCS cache data \cite{cache_data}.
This can also leak the sensitive data, if privacy data running in enclave located in malicious party.

\textbf{SRAS negotiates a common TCB platform information, isolated from privacy enclave TEEs.} 
Since the trustworth of all parties does not leave the local party, we have another unique key observation:
{\itshape if all TCB platform information in each relying party are a consensus negotiated by all parties, 
and all TEE verification occurred in the relying party abides by this consensus, the TEE verification will not be comprised.}
Therefore, we design a $policy$ strategy to record all RPE TCB platform information, privacy enclave’s TCB out of data information and identity digests, etc.
When RPE does mutually attestation and locally privacy enclave attestation, the RPE does verification based on the TCB information recorded in $policy$ 
(in other words, RPE verification does not need PCCS to provide TCB information).
However, a new challenge for us is how to prevent the consensus $policy$ from being tampered with before it is configured into the RPE memory.
Fortunately, if we extent $policy$ hash into the DCAP Quote, then each RPE can compare consistency of hash value and its stored $policy$ when mutual attestation.
Because RPE's behaviors are publicly verified and consistent per party, all verifications performed within the RPE strictly follow the $policy$ strategy.
For the privacy enclave's TEE, we use a TCB out of data list in $policy$ to prevent privacy enclave from being launched in a vulnerable environment. 
But other participants do not know the privacy enclave's specific platform information.

\begin{figure}
	\centering
	\includegraphics[width=0.475\textwidth]{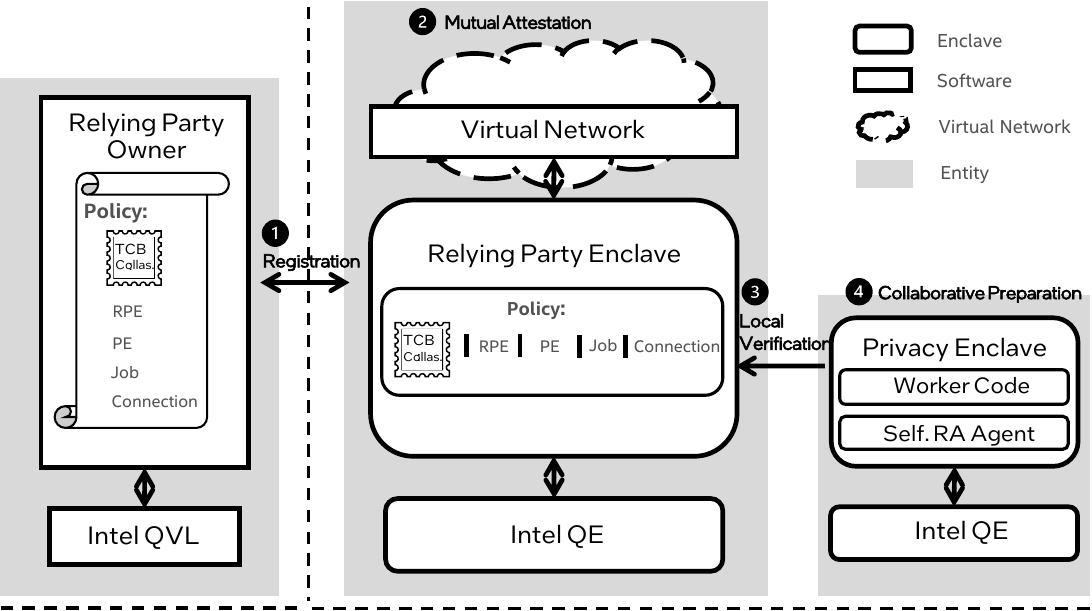}
	\caption{SRAS microservice architecture.}
	\label{fig:overview}
\end{figure}

\subsection{Overview}
From the insights above, we have created SRAS, a self-governed remote attestation scheme.
It does verification for each local privacy enclaves and provides a secure communication channel for all collaboration parties.
As illustrated in Figure \ref{fig:overview}, it contains four components:
\begin{itemize}
\item Relying Party Owner (RPO). 
It is the trust anchor of each participant and located in intranet world for each party.
It is responsible for attesting locally RPE based on Intel Quote Verification Library (QVL) and pass the consensus $policy$ to the RPE.  
 
\item Relying Party Enclave (RPE).
It inherits the trust anchor and attests locally privacy enclaves for each party.
RPEs conduct the mutual remote attestation based on $policy$ and verify all verification messages come from other parties.
It also generates a pair of signing keys, which serve as identifiers for each participant and are used to sign or verify all messages sent or received.

\item Virtual Network.
It is a network agent that need to configure for all parties in advance. 
It receives messages from RPE and send to the other parties.
It also documents all received messages for logging purposes. 

\item Privacy Enclave (PE). 
It contains the intellectual property enclaves for each party (called Work Code). 
It is responsible for communication with RPE (called Self.RA Agent). 
It also generates the keys that are used to build a secure collaboration channel.
\end{itemize}
Therefore, SRAS consists of four phases:


\begin{lstlisting}[frame=single,
                label={policy_example}, caption={$Policy$ strategy example.}
               framesep=0mm,
               xleftmargin=0pt,
               numbersep=0pt]{js}
{ "Session ID": "uuid",

  "TCB": 
  [  {"id": "tcb-1",
      "fmspc": "fmspc value 1",
      "data": "collateral"},
     {"id": "tcb-2",
      "fmspc": "fmspc value 2",
      "data": "collateral"}],
      "Out of Data TCB": 
  [  {"id": "tcb-1",
      "fmspc": "fmspc value 1",
      "data" : "collateral"}],
  "RPE": 
  [  {"entity": "rpe-1",
      "qeid_allowed":["qeid1", "qeid2"],
      "tcb_allowed" : ["tcb-1"]},
     {"entity": "rpe-2",
      "qeid_allowed": ["qeid3"],
      "tcb_allowed" : 
          ["tcb-1", "tcb-2"]}], 
  "PE": 
  [  {"entity": "pe-1",
      "mrenclave": "mrenclave",
      "mrsigner_allow_any" : true,
      "isvprodid_allow_any": true,
      "isvsvn_allow_any"   : true},
     {"entity": "pe-2",
      "mrenclave_allow_any": true,
      "mrsigner" : "mrsigner",
      "isvprodid": 0,
      "isvsvn_minimum": 0}],
  "Job": 
  [  {"id": "job-1",
      "rpe": "rpe-1",
      "pe" : "pe-1",
      "pe_qeid_allowed": ["qeid1"],
      "out_of_tcb"     : ["tcb-1"]},
     {"id": "job-2",
      "rpe": "rpe-2",
      "pe" : "pe-2",
      "pe_qeid_allowed": ["qeid2"],
      "out_of_tcb"     : ["tcb-1"]}],
  "Connection": 
  [  {"server" : "job-2",
      "clients": ["job-1"]}]
}
\end{lstlisting}

\noindent \textbf{\itshape Registration.} 
We firstly customize the strategy $policy$ based on the consensus of all parties involved.
All subsequent validations by all parties will be based on this $policy$.
In this phase, the trust anchor of each participant is RPO.
RPO then validates the legality of locally RPE based on RPE identity negotiated in $policy$ strategy.
If attestation successfully, RPO passes the $policy$ to RPE, and RPE will be unique representative of each participant for attestation and verification. 
In other words, RPE inherits trust anchor from RPO in each party.

\noindent \textbf{\itshape Mutual Attestation.} 
After received $policy$ strategy, RPE performs mutul attestation, forming the virtual verifiable network to communicate the verification messages.
Each RPE first verifies the correctness of $policy$ hash value and other participants' RPE identity based on received others' RPE Quotes, 
preventing to launch a malicious RPE or tamper the $policy$ strategy.
If validation successfully, each RPE records the public signing key generated by other RPE embedded in the Quotes as other RPEs' identifier.  
RPE uses its recorded public keys to verify messages passed in the virtual network.  

\noindent \textbf{\itshape Local Verification.} 
According to the $policy$ strategy each RPE performs a locally verification for local PE.
If PE's quote is validly verified and its identity is match recorded in $policy$, 
each RPE updates the verification results signed by itself to the virtual network. 
Each RPE then validates the verification results by using recorded other parties' RPE public key.
Once verification results of other parties' PEs are verified successfully, 
The PEs generate the public key certificates and are signed by local RPE.
They will be used as identity verification when building secure channels for multi-party collaboration. 
At this point, the verification results of each participant's local PE have been accepted by the other participants.

\noindent \textbf{\itshape Collaborative Preparation.} 
In this phase, PE takes the collaborative preparation for building secure channel.
According to the collaborative rules in the $policy$ strategy, 
each PE initiates a connection with the enclave it wants to collaborate with to exchange identity information (public key certificates).
PE sends received identity message to the local RPE for verification.
RPE uses recorded other parties' public key to validate the identity since these information has been signed by RPE. 
If verification successfully, PE will use the counterpart identity public key to build secure channel for collaboration.


\section{Detailed Design}\label{design}
\subsection{Registration}\label{registration}
In this phase, each participant negotiates the collaborative consensus to form a $policy$ strategy.
Each participant attests local RPE based on this collaborative $policy$.
Then, trust anchor is passed to RPE.

\noindent \textbf{Negotiating the consensus.} 
In order to achieve the same behavior of local attestation for all parties, i.e., the local RPE can represent other parties, 
and also to address the TEE configuration challenge (C3), we extracted the key information used for multi-party attestation to form a consensus.
As shown in the List \ref{policy_example}.

\begin{itemize}
\item \verb|Session ID| defines a identifier, which uniquely identifies the identity of the current round of collaboration. 
If this collaboration is updated or a new collaboration is launched at the end of this round, then the identifier will also change. 

\item \verb|TCB| identifies all TCB hardware platform information used to run the parties' RPEs.
The \verb|fmspc| \cite{cache_data, pck} defines a description of the processor package or platform instance. 
The \verb|data| defines a TCB information for the given platform.

\item \verb|Out of Date TCB| identifies all TCB hardware platform information that has expired.
Include some hardware platforms with vulnerabilities.

\item \verb|RPE| identifies the specific hardware platform (\verb|qeid_allowed| embedded in the Quote and generated by QE\cite{dcap, cache_data}) on which the RPE is allowed to launch for each party, 
as well as containing TCB platform information for this platform.

\item \verb|PE| identifies the launched privacy enclaves' identity for each party, including anonymized information about \verb|mrenclave|, \verb|mrsigner|, and software products such as \verb|isvprodid|.

\item \verb|Job| defines attestation relationship of local RPE and PE for each party.
It also defines the specific hardware platform that is able to launch PE in \verb|pe_qeid_allowed|, 
and outdated TCB information for the given platform.

\item \verb|Connection| defines a collaborative relationship for all parties.
\end{itemize}

\begin{figure}
	\centering
	\includegraphics[width=0.4\textwidth]{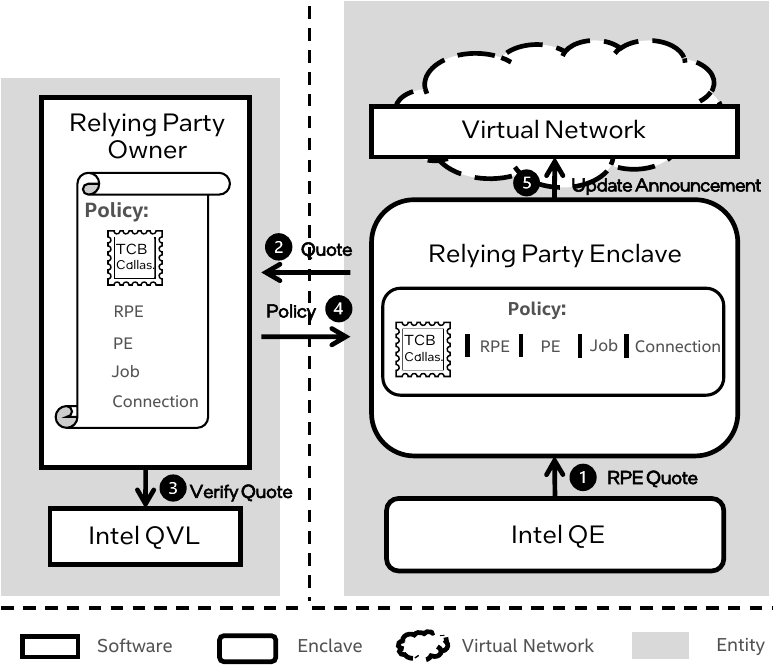}
	\caption{Registration Phase.}
	\label{fig:phase1}
\end{figure}

\noindent \textbf{Attesting RPE.} 
After negotiating $policy$ strategy, 
each participant' RPO verifies their own local RPE to complete the trust anchor conversion.
As shown in Figure \ref{fig:phase1}, take one of the participants as an example.
RPE communicates with the QE to generate RPE Quote and send it to RPO located in intranet world (step 1 and 2).
Then RPO verifies the Quote based on the Intel QVL.
Based on the RPE identity recorded in $policy$ (\verb|RPE| described in List \ref{policy_example}), 
RPO verifies that the RPE's TEE is secure (step 3).
If verifying successfully, RPO will send the $policy$ to RPE (step 4).
At this point, the trust anchor of the participants is transferred to the RPE.
Finally, RPE generates a pair of signing public and private keys.
In step 5, RPE signs an announcement and updates to the multi-party virtual network, announcing its entry into the next phase.

We can potentially eliminate the privacy data of each participant from being leaked to others.
In $policy$ strategy, all identities of privacy enclaves are anonymously documented.
Since all PEs are attested in the enclave of the RPE located locally, 
including that the RPE receives PE quotes and compares their identity with the information in the recorded consensus. 
All of these processes exclude other parties.

Additionally, 
\verb|TCB| is documented in the policy to prevent local TCB information from being replaced and thereby authenticating the RPE launched in an unexpected TEE (Quote verification details are described in \ref{bkg:DCAP}).
\verb|Out of Data TCB| also prevents PE from being launched in a vulnerable TEE. 
It ensures that an attacker cannot snoop on secrets in PE. 
Also, we define the \verb|qeid| allowed list for RPE and PE preventing rogue RPE or PE from being launched to get collaborative data.
Because the hardware platforms approved by all parties have been bound in the $policy$, and they will be verified during attestation. 

\begin{figure}[htbp]
	\centering
	\includegraphics[width=0.4\textwidth]{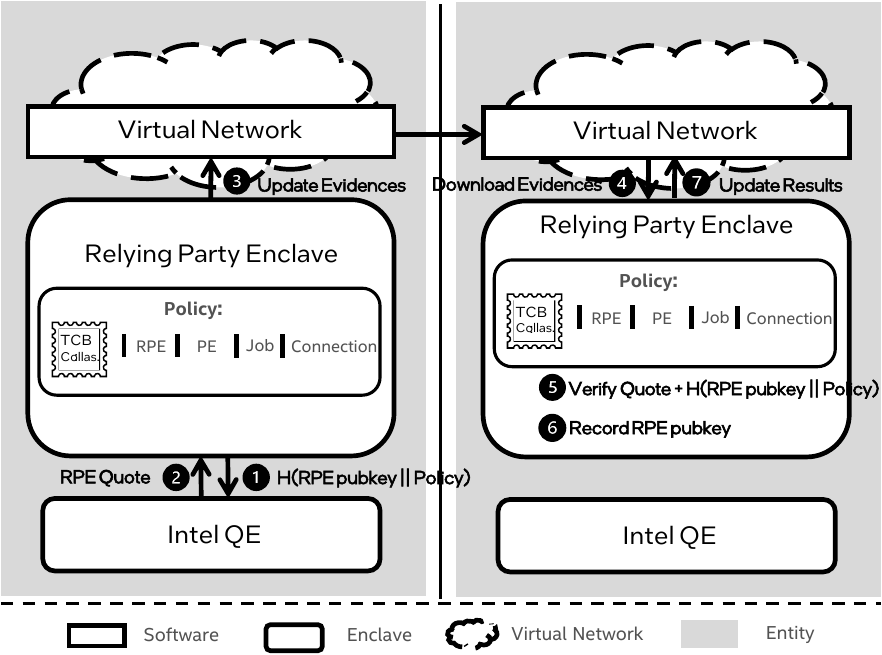}
	\caption{Mutual Attestation Phase. Assuming that there are two participants in multi-party cloud computing. }
	\label{fig:phase2}
\end{figure}

\subsection{Mutual Attestation}
Once the RPE has been successfully validated, the second phase starts. 
RPEs conduct the mutual remote attestation based on negotiated $policy$, and form a virtual verifiable network.

\noindent \textbf{Generating RPE evidences.}
In Figure \ref{fig:phase2}, RPE hashes its public signing key and $policy$ into Quote.
It fill the hash valve into $report\_data$ field of Quote Enclave Report (step 1 and 2).
Then RPE generates the RPE evidences and update them to the Virtual Network component.
The Virtual Network component broadcasts these evidences to other parties for validation (step 3).
Also it receives evidences from other parties for the purpose of RPE's authentication of other parties. 

\begin{table}[h]
    \centering
    \caption{RPE Evidence Structure. }
    \label{tab:phase1}
    \scalebox{0.8}{
    \begin{tabular}{|l|p{0.68\columnwidth}|}
         \hline
         \bfseries Item & \bfseries Description \\
         \hline
         Entity &  Participant represented by RPE.\\
         \hline
         Identifier & Public signing key generated by RPE.  \\
         \hline
         Quote & Enclave Quote generated by RPE.\\
         \hline
         Consensus RA Results & Whether the other party's $policy$ strategy is accepted by own party or not.\\
         \hline
    \end{tabular}}
\end{table}

Evidences structure is shown in Table \ref{tab:phase1}.
The Virtual Network component maintains all evidence records in each party for this phase.
Note that the \verb|Consensus RA Results| item is \verb|no| at this point.
This value will not be set to \verb|yes| until all other evidences have been verified by local party's RPE.

\noindent \textbf{RPE mutual attestation.}
After the Virtual Network component of each party received the other parties' evidences, it will send them to local RPE for verification.
The Virtual Network component on the right side of Figure \ref{fig:phase2} receives the counterpart's evidence and send it to local RPE (step 4).
RPE fetches the counterpart Quote and \verb|Entity| from received evidence structure.
With the \verb|Entity| key, RPE retrieves RPE identities in $policy$ strategy and obtains the corresponding RPE TEE description.
Based on this TEE hardware platform description, RPE verifies the validity of counterpart RPE Quote structure.
Besides, the RPE still verifies the validity of three other items:
\begin{itemize}
\item RPE first verifies that the other party's $policy$ strategy is consistent with the local's by comparing $policy$ hash in $report\_data$ with its local $policy$ hash. 
Ensuring that all participants follow the negotiated consensus to local attesting.
\item RPE verifies that the other party's identifiers are valid by comparing RPE signing key hash with the identifier value in the received other party's evidence structure.
\item RPE verify that the other party's TEE hardware platform on which the RPE is launched is the pre-specified platform by extracting the \verb|qeid| from received RPE Quote and comparing the value with the corresponding \verb|qeid_allowed| in local $policy$ strategy.
\end{itemize}

After validating the above three items (step 5), 
RPE considers that the local RPEs of other participants have been effectively configured and launched normally.
Each party's RPE performs local attestation based on the pre-agreed consensus strategy.
In other words, each party's RPE can perform local authentication on behalf of the other parties.
Note that the RPE measurements are also verified implicitly during mutual attestation.
Because the measurements of all RPE are the same.

\noindent \textbf{Recording RPE identifiers.}
In step 6, RPE stores the verified other RPE's identifier.
At this point, the RPEs of all participants have completed the mutual attestation.
The virtual verifiable network formally established, 
in which each RPE uses recorded other RPE's public key identifier to verify the messages received from other parties for multi-party collaboration.  
Finally, each RPE signs a \verb|yes| message and updates to the \verb|Consensus RA Results| item of its own evidence structure and broadcasts it to the others, 
announcing entry into the next phase (step 7).

\begin{figure}
	\centering
	\includegraphics[width=0.4\textwidth]{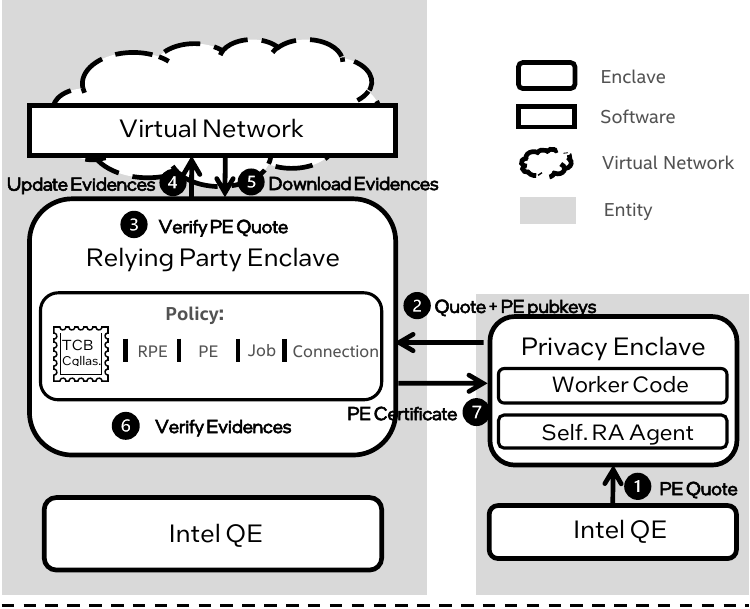}
	\caption{Local Verification Phase. }
	\label{fig:phase3}
\end{figure}

\subsection{Local Verification}
When all \verb|Consensus RA Results| items are set to \verb|yes|, RPE begins attesting the locally privacy enclaves. 
After successful attestation, each RPE communicates the PE attestation results with each other,
enables acceptance of local results by all parties. 

\noindent \textbf{Attesting PE.}
As illustrate in Figure \ref{fig:phase3}, PE communicates with QE to generate PE Quote (step 1).
At the same time, PE also generates a pair of signing keys used for subsequent building of the secure channel.
PE sends the Quote and PE public key to the local RPE (step 2).
According to the \verb|Job| relationship in the $policy$ in List \ref{policy_example}, 
the local RPE retrieves and obtains the anonymized measurements of the PE for which it wants to be attested, and attests the PE (step 3).

For example, local RPE entity is \verb|rpe-1| and \verb|Job| relationship is \verb|job-1|.
According to the relationship, 
the RPE retrieves that the PE to which it wants to be attested is the entity \verb|pe-1|.
Therefore, local RPE continue to retrieve and obtain the anonymized measurements of \verb|pe-1|.
Based on the measurements of PE, RPE verifies the validity of received \verb|pe-1| Quote.
Besides, local RPE also verifies that the \verb|pe-1|'s TEE hardware platform on which PE is launched is the pre-specified platform.  
It extracts the \verb|qeid| from \verb|pe-1| Quote and compares the hash value with the corresponding \verb|pe_qeid_allowed|. 
If the hardware platform is the same as pre-specified, the RPE will continue to verify that the TCB for the verified hardware platform is not out of date. 
RPE compares the TCB information in \verb|pe-1|'s Quote with the value described in \verb|out_of_tcb|, guaranteeing that \verb|pe-1|'s TEE is not vulnerable.  

\begin{table}[h]
    \centering
    \caption{PE Evidence Structure. }
    \label{tab:phase3}
    \scalebox{0.8}{
    \begin{tabular}{|l|p{0.68\columnwidth}|}
         \hline
         \bfseries Item & \bfseries Description \\
         \hline
         Entity & Participant-launched privacy enclaves.\\
         \hline
         PE RA Results & The result of the RPE attesting the PE and signed by the RPE.\\
         \hline
    \end{tabular}}
\end{table}

\noindent \textbf{Verifying other parties' attestation results.}
After verifying the PE Quote, each RPE generates PE evidences and updates them to the Virtual Network component, 
broadcasting its evidence to other parties to enable them to validate the results of the local PE attestation.
The PE evidence structure is shown in Table \ref{tab:phase3}.
The RPE signs the PE RA Results using its signing key generated in Sec. \ref{registration} and updates the PE evidence to other parties (step 4). 
RPE then retrieves the \verb|Connection| relationship described by the $policy$ in List \ref{policy_example} to find the participants who will be collaborating.
RPE downloads the evidence of the PE to be collaborated through the virtual verifiable network for validation (step 5).

Continuing the example above, the Job relationship is \verb|job-1|.
According to the \verb|Connection| relationship, 
the collaborative target of the \verb|pe-1| in \verb|job-1| is \verb|pe-2| which is located in \verb|job-2|.  
Therefore, local RPE \verb|rpe-1| downloads the PE evidence updated by the \verb|rpe-2|.
Local RPE \verb|rpe-1| uses the \verb|rpe-2| public key identifier recorded during the Mutual Attestation phase to verify the validity of the PE attestation result in the PE evidence of \verb|pe-2| (step 6). 
If the validation is successful, local RPE \verb|rpe-1| considers that the privacy enclave \verb|pe-2| has been validly verified by its local RPE \verb|rpe-2|.
In other words, \verb|rpe-2| successfully performs local attestation on behalf of \verb|rpe-1|.

\begin{table}[h]
    \centering
    \caption{Structure of PE certificate. }
    \label{tab:PEcert}
    \scalebox{0.8}{
    \begin{tabular}{|l|p{0.58\columnwidth}|}
         \hline
         \bfseries Name & \bfseries Description \\
         \hline
         PE public key & PE public key for building secure channel.\\
         \hline
         Session ID  & Retrieve from the $policy$ consensus, valid only for $policy$-limited collaboration cycles.\\
         \hline
         Nonce & A nonce generated within RPE.\\
         \hline
         RPE verification report & RPE signature structure whose report data is the hash of the three entries above.\\
         \hline
    \end{tabular}}
\end{table}

\noindent \textbf{Issuing PE certificates.}
When the RPE of each participant completes the results verification of all the PE evidence to be collaborated, 
they sign the PE public key received in step 2 to form the PE certificate and issue it to the local PE (step 7).  
The structure of the PE certificate is shown in Table \ref{tab:PEcert}.
Note that there is no strict timestamp restriction in the PE certificate because the Session ID in $policy$ only applies to the current collaboration.
It will be updated if the collaboration changes.
PE certificates will be used to build a secure channel for the next phase of collaboration.

\subsection{Collaborative Preparation}
Since the local verification results have been accepted by other parties, 
there is no need for remote attestation between the privacy enclaves that are collaborating.
They build secure channel directly by verifying each other's PE certificates.

\begin{figure}
	\centering
	\includegraphics[width=0.45\textwidth]{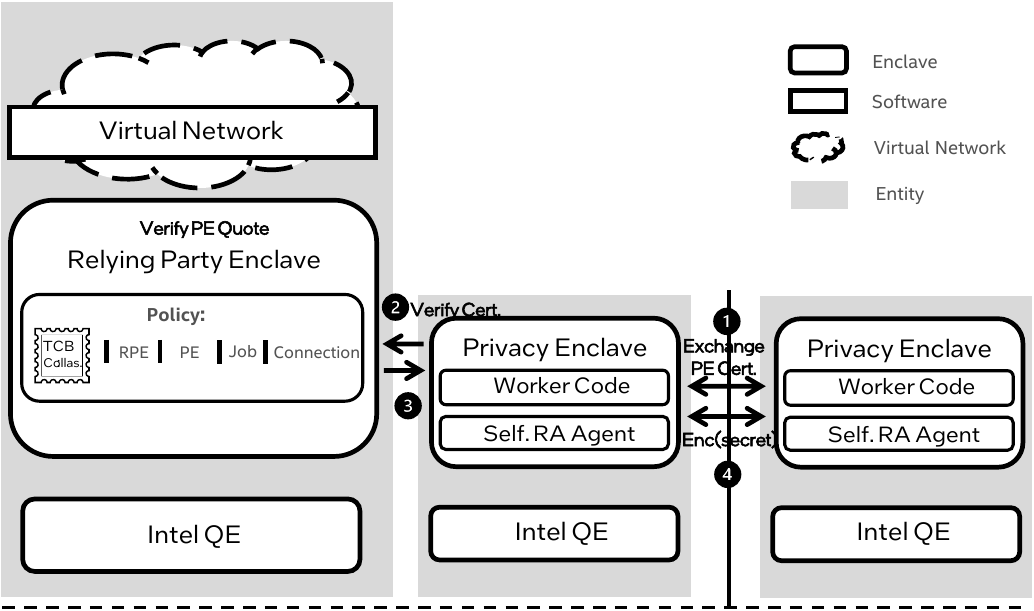}
	\caption{Collaborative Preparation Phase. }
	\label{fig:phase4}
\end{figure}
\begin{figure}
	\centering
	\includegraphics[width=0.485\textwidth]{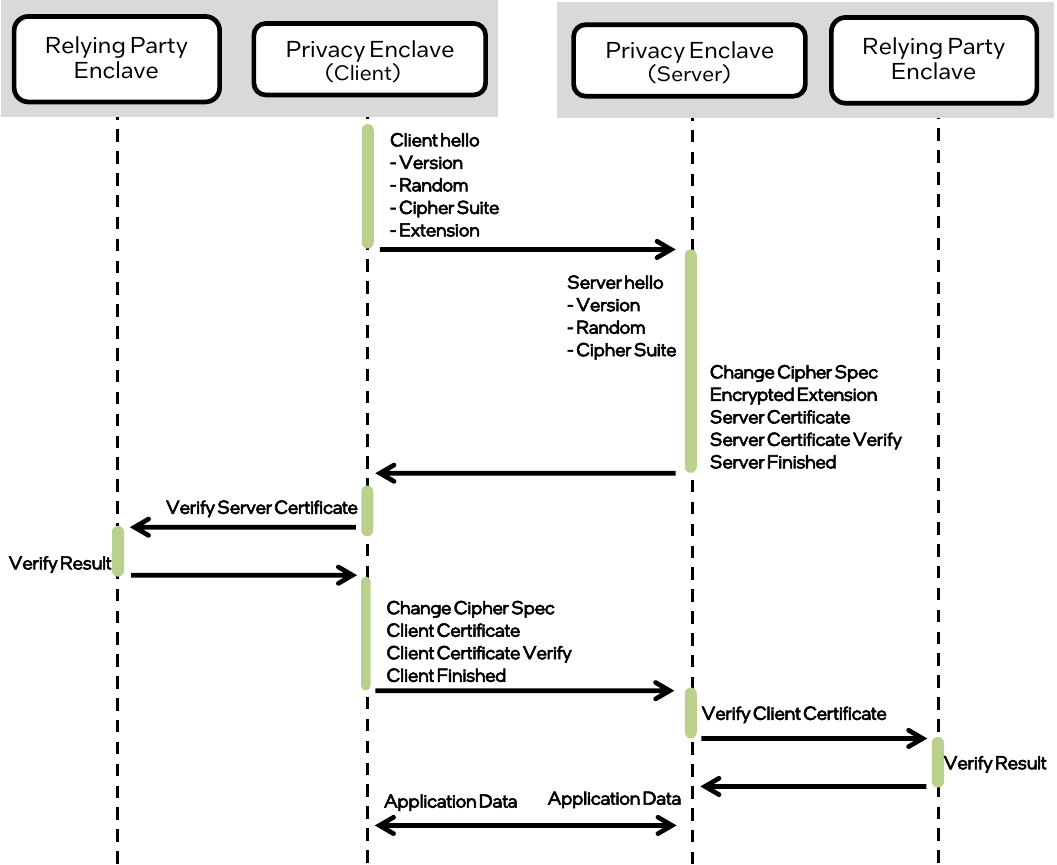}
	\caption{Two-way TLS 1.3 Handshake. }
	\label{fig:4case}
\end{figure}

\noindent \textbf{Exchange and Verify PE certificates.}
In Figure \ref{fig:phase4}, when the PE communicates with the PE it wants to collaborate with, 
it sends the PE certificate of the other party to the local RPE after receiving it to verify the validity of the PE certificate (step 1 and 2).
Based on the \verb|Connection| relationship in the $policy$, 
the local RPE looks up the public key identifier of the recorded RPE, 
which is generated by the local RPE of the collaborated PE (e.g., \verb|rpe-2| in the above example). 
The RPE uses the recorded RPE identifier to verify the received PE certificate.
If verification is successful, the RPE will notify the local PE that it can securely send application data with the collaborating PE (step 3 and 4).

\noindent \textbf{Secure Channel Example.}
We describe a Transport Layer Security (TLS) example to demonstrate how to build a secure channel between two collaborating PEs using RPE verification.
The workflow is illustrated in Figure \ref{fig:4case}.

The client PE (1) sends a Clienthello messages to server PE. Server processes the Clienthello messages and responds with its own Serverhello.
Server then (2) sends the EncryptedExtension and Server Certificate to client. 
This certificate is issued by its RPE.
Client (3) communicates with its local RPE to verify the Server Certificate.
Client then (4) responds with its own Client Certificate.
Server (5) verifies the Client Certificate by communicating with its local RPE.
Collaborating parties can now securely send application data to each other.

\section{Implementation}\label{implementation}



Our implemented SRAS consists of four components: the Relying Party Owner (RPO), the Relying Party Enclave (RPE) including some help modules, 
the Privacy Enclave (PE), and the Blockchain which is for RPEs to build a Virtual Network with each other. 
In the implementation of our prototype, we adopted three technologies: Gramine LibOS \cite{gramine}, RA-TLS protocol \cite{knauth2018integrating} and Hyperledger Fabric \cite{fabric}. 
We adopted Gramine (version 1.4) to implement the RPE and PE, which is released under the GNU Lesser General Public License version 3.0. 
We adopted Hyperledger Fabric (version 1.4) to implement the Virtual Network, which is released under the Apache License version 2.0. 
We use the RA-TLS protocol for communication between RPO and RPE as well as between RPE and PE.

\noindent \textbf{Relying Party Owner} is responsible for invoking Intel QVL using \verb|verifyRPE()| to attest and verify the local RPE,
and pass the policies to RPE by \verb|LoadPolicy()|.

\noindent \textbf{Relying Party Enclave} is responsible for attesting and verifying the local CE, 
and building the Virtual Network between other parties to exchange information with each other. 
There are several functions implemented by RPE: parse $policy$, generate RPE evidence, attest and verify counter-part RPEs’ or PE’s evidence, synchronize information to blockchain, issue and verify PE certificate.

\verb|ParsePolicies()| parses the $policy$, which is reached agreement by relying parties and received from RPO, to three components: RPE’s measurement and TCB information, local PE’s measurement and TCB information, and PEs’ connection relationship. 
The $policy$ including information required for all stages, and only parsed them correctly can RPE execute the upcoming steps according to the agreed rules.

\verb|GenerateEvidence()| generates the evidence for RPEs to attest and verify each other, with the hash of RPE’s public key and $policy$ set in $report\_data$, which is a structure in evidence. 
And the hash will be verified by other RPEs when attesting each other’s evidence. 
This is one of the key design in the SRAS, which guarantees that all the relying parties will follow the same $policy$. 
The steps of generate evidence are: generate report including measurements and sign report to form the certificate chain and evidence. 
These two steps are implemented by Gramine, and we can only provide the report data and read a special file “/dev/attestation/quote” in Gramine to get the evidence.

\verb|VerifyEvidence()| attests and verifies the evidence got from other RPEs or the local PE. 
The collaterals required such as certificate revoke lists and TCB information for attestation is read from $policy$. 
When verifying counter-part RPEs’ evidence, the measurements, and the hash of the public key and $policy$ will be checked, in which the measurements are read from the local $policy$, the public key is transformed along with evidence, and the $policy$ is read from the local. 
When attesting and verifying the local PE’s quote, the function is integrated into a Gramine library called RA-TLS, which provide Remote Attestation via TLS connection.  

\verb|SynchronizeInformation()| provides RPE’s helper procedure the ability to connect to the Blockchain. 
RPE with permission to the network can upload information to or download information from the Blockchain when calling it. 
This function is for RPEs to synchronize information with each other, such as, RPE’s quote, public keys, RPE’s verification signature and the endorsement for PEs.

\verb|IssuePECertificate()| is designed for the RPE to issue a certificate to the local PE. 
This certificate is generated using the RPE's signing private key after the completion of the verification process. 
It serves as an endorsement for the PE and is utilized by the PE to establish a connection with the collaborative one.

\verb|VerifyPECertificate()| enables the RPE to verify the certificate of the counterpart PE. 
This verification process involves using the signing public key of the corresponding RPE. 
When two PEs are prepared to establish a connection, it is essential to complete this verification step. 
Successful verification is a prerequisite for PEs to connect with each other and engage in subsequent collaboration activities.

\noindent \textbf{Privacy Enclave} is responsible for running user applications and providing the functions for several applications with collaborative relationships. 
We implement the basic functions such as evidence generation, certificate getting, data exchange, etc, for ISVs to complete the development of PE and use it in actual scenarios.

\verb|GenerateEvidence()| generates the evidence quote for PE to be verified by the local RPE, with the hash of PE’s public key set in $report\_data$. 
And the hash will be verified by the local RPE when attesting the evidence quote. 
The steps of generate evidence are the same as RPE’s and are also implemented by Gramine.

\verb|GetPECert()| serves the purpose of a PE obtaining the certificate of its counterpart PE during the TLS connection establishment process. 
Upon obtaining the certificate, the PE proceeds to transmit it to the local RPE for signature verification. 
If the verification process is successful, indicating the integrity of the certificate, the connection can be established successfully.

\verb|PEExchangeData()| is designed for two PEs to securely exchange information with each other after successfully establishing a TLS connection. 
This mechanism ensures a secure channel for the exchange of data between the two PEs within the TLS connection, enhancing the confidentiality and integrity of the transmitted information.

\noindent \textbf{Blockchain}.
The responsibility of blockchain is to play the role of the Virtual Network for the relying parties. 
Relying parties can exchange information like evidence through the blockchain. 
There are some functions in this blockchain: register, query and update enclave information including RPE and PE. 
We use smart contracts in Hyperledger Fabric to implement these functions. 
That why SRAS design separates out the Virtual Network component separately. 

\section{Analysis and Evaluation}\label{evaluation}

\subsection{Security Analysis}
The goal of SRAS is to provide a Relying Party with attestation and verification functions to validate the trustworthiness of TEE and compute assets.
Our security analysis focuses on attackers substituting or forging Relying Party for attestation and verification.

\noindent \textbf{Provide forged policy to RPE.} 
If the RP owner is malicious, or if the communication channel between the RPO and the RPE is compromised so that a forged $policy$ is passed to the RPE, 
SRAS is able to identify these attacks and refuse to perform RA attestation.
Because during the RPE mutual attestation process, each party's RPE will verify whether the other party's $policy$ hash is the same as its own $policy$.
The passing of the $policy$ hash is endorsed by the RPE Quote, and the $policy$ must remain the same across all participants.
Besides, if a replay attack is performed using an old $policy$, the Session ID of the $policy$ will block such an attack.
Because the Session ID for each collaborative policy is different.

\noindent \textbf{Compromise the verifiable network to modify the evidences.} 
If the verifiable Virtual Network is attacked, for example if a malicious blockchain service deletes or alters the data for the chain, 
the integrity of the evidences is compromised, the RPE refuses to operate, 
and SRAS notifies that human intervention is necessary to restore operation.
Because the evidence passed to the verifiable network have been signed by its own RPE.
The RPE public keys of other parties recorded by each party have been endorsed by the RPE Quote during the RPE mutual authentication process.
As a result, the RPE is able to verify the integrity of the received evidence by verifying the signature of the evidence with other PRE public keys on record.
In addition, if an attacker utilizes old evidence for a replay attack, e.g., during the mutual or local attestation phase, 
other RPEs will not be able to verify the evidences.
Because the RPE public keys associated with evidences are different for each collaboration.
And the RPE private key never leaves the enclave memory.

\subsection{Performance Evaluation}
We evaluate the performance of our prototype implementation of SRAS.  
All experiments were run on machine with Intel Xeon Platinum 8352Y CPU running at 2.20GHz, 16 GB RAM and 8 logical cores.
Our machine runs Ubuntu 20.04 x86\_64 GNU/Linux.  
Since the performance results are highly correlated with the execution time of each phase,
we evaluate the latency of different phases.

\noindent\textbf{Registration Latency.} 
We first measure the latency during the Registration phase.
We run the Registration phase 10000 times and calculate the average.
We report a latency of 121.1 ms for the Registration phase.
 
\noindent\textbf{Mutual Attestation and Local Verification Latency.} 
We measure the latency of Mutual Attestation and Local Verification phases.
Since both phases download evidences from the Virtual Network, 
and the Virtual Network is a network agent plug-in in SRAS and can be replaced by other third-party agents (in our implementation we use Fabric blockchain), 
the latency downloading evidence from other parties depends on the implementation of the network agent plugin, 
which is out of our control. 
Therefore, we ignore the waiting latency of downloading the evidences and only focus on the latency of processing the evidences in the SRAS softwares.
In Mutual Attestation phase, 
we report the latency from generating RPE Quote to updating RPE evidence (step 1 to 3, in Figure \ref{fig:phase2}) is 129.2ms (average over 10000 runs).
We report the latency from downloading RPE evidences to recording RPE public keys (step 4 to 6) is 53.5ms (average time to process 10000 pieces of evidence).
In Local Verification phase,
we report the latency from generating PE Quote to updating PE evidence (step 1 to 4, in Figure \ref{fig:phase3}) is 193.7ms (10000 runs).
We report the latency from downloading PE evidences to issuing PE certificate (step 5 to 7) is 2.8ms (process 10000 pieces of evidence).

\noindent\textbf{Collaborative Preparation Latency.} 
We first measure the latency of building secure channel for two PEs in Collaborative Preparation.
We report the latency is 87.6ms (average over of 10000 runs).
We then report the average latency is 6.3ms for verifying the PE certificate of the counterpart (step 2 to 3, in Figure \ref{fig:phase4}).
The latency in verifying certificates is approximately 7\% of the total latency in the Collaborative Preparation phase.

We summary all latency for all phases in Table \ref{tab:latency}.
{\itshape Before} indicates the latency before uploading to the Virtual Network.
{\itshape After} indicates the latency after downloading from the Virtual Network.
Since the Registration phase does not download any information from the Virtual Network, the {\itshape After} item is denoted as N/A.
We observe that the total latency for all phases is much less than 0.5 seconds.
Once attestation and verification are completed, the secure channel construction time of 87.6ms is acceptable for the next step of collaboration.

\begin{table}[h]
    \centering
    \caption{Total Latency.}
    \label{tab:latency}
    \scalebox{0.8}{
    \begin{tabular}{|l|c|c|c|}
    \hline
    \multicolumn{1}{|c|}{\multirow{2}{*}{Phase}} & \multicolumn{3}{c|}{Latency/ms}\\ \cline{2-4} 
    \multicolumn{1}{|c|}{} & 
    \multicolumn{1}{c|}{Before } & 
    \multicolumn{1}{c|}{After } & Total latency \\ \hline
                       Registration &       121.1 &    N/A &   121.1\\ \hline
                       Mutual Attestation &   129.2 &   53.5 & 182.7 \\ \hline
                       Local Verification &   193.7 &   2.8 & 196.5 \\ \hline
                       Collaborative Preparation &  N/A & N/A & 87.6 \\ \hline
    \end{tabular}}
\end{table}

\section{Discussion}\label{discussion}

\noindent\textbf{Security product upgrade issues.} 
In our design, 
we are able to address the overhead issues of frequent upgrades of security products that require reconfiguration and verification in a ready-to-trust workloads.
To prevent exploitation of potential vulnerabilities, TEE manufacturers usually provide post-product security updates for supported product,
such as Intel's TCB upgrade for SGX TEE.
However, some users may reject security upgrades because they can not afford the time overhead of relaunching workloads or because the new vulnerabilities are within their risk tolerance.
SRAS allows parties to use different versions of the TCB platform.
Collaboration can continue as long as other participants accept the current TEEs.
Because all participants' TEEs are verified based on the TCB information bound to the policy in the RPE.


\noindent\textbf{Independent Third Party support.}
Although SRAS is a decentralized relying party system, 
it still seamlessly supports attestation plans and notary services provided by independent Third Party.
The Third Party may first review the RPE software and then publicize and distribute the official version of the RPE, 
including the RPE measurements.
The end user then registers their RPO software with the Third Party.
After authorization, RPO generates a deployment private key from a key generator plug-in provided by the Third Party, 
whose corresponding certificate is filed by the independent Third Party.
When the RPE registers with the RPO, 
the RPO passes the deployment private key as secret to the RPE through the established secure channel.
The RPE then uses this deployment privacy key to sign the attestation results of the RPE attesting the PE.
When users want to verify the attestation result of the PE, 
they can verify the signed attestation result by querying the deployment key certificate from the independent Third Party.
For scenarios where data owners want to put private data into PE for calculation, 
they can obtain PE certification results from RPE and validate them with the independent Third Party recognized by them. 

\noindent\textbf{Logging support.}
In our design, SRAS also supports logging.
There are two log output strategies in the SRAS. 
{\itshape RPO attestation log:} 
Recording the RA results for RPO attesting RPE (The results are also uploaded to the virtual network for validation by other participants.).
{\itshape RPE attestation log:}   
1) RPE Evidence Table records the RPE identities and RA results of RPEs mutual attestation.     
2) PE Evidence Table records the RA results for RPE attesting PE. 
However, how to prevent logging information to compromise when a user query the logs?
In SRAS, we can set a certificate field called \verb|signing_key_cert| in $policy$. 
This field is used for each RPO user to get a certificate from a CA authority that is recognized by all. 
All RPO users fill in their certificate into $policy$, and upload $policy$ to its own RPE, together with the corresponding signing key.
By this, RPO users use the CA private key to sign the RA results for RPO attesting RPE. 
Therefore, the validators can use the public key certificate filled in the $policy$ to verify the RA results signature.
Furthermore, the signed RA results in {\itshape RPO attestation log} also include the RPE public key.
Since SRAS uses the RPE private key to sign all Evidences in the virtual network, 
the validator can verify the integrity of all Evidences using the RPE public key obtained from the RA results in the {\itshape RPO attestation log}.


\section{Related Work}\label{relatedwork}
Existing researches are also making efforts to address the challenges of attestation and verification for multi-party TEE,
but only partially.
MAGE \cite{chen2022mage} proposes to split the enclave into two parts, 
one retaining the original functional logic code and the other that derives the identity of the multi-party enclave for attestation.
It makes groups of enclaves mutually attestation without third-party relying party, 
but still requires centralized development and distribution.
In addition, it also does not support enclave updates.
Once an existing party's enclave is modified, all parties' enclaves within the group need to be modified.
SRAS does not have such issues because of its ready-made Relying Party design.
Once one party's enclave update, each Relying Party of each party need only re-attestation according to its in-process policy.
Greveler et al. \cite{greveler2012mutual} propose a mutual remote attestation protocol to verify the secure cloning of trusted platform.
They consider that the TEE must be the identical for all platforms.
However, SRAS is not bound by this limitation and each party can have a different TEE.
Apache Teaclave \cite{Apache} addresses multiple enclaves attestation by relying on third-party auditors. 
But it is very difficult to find a reliable third party in a multi-party scenario.

OPERA \cite{chen2019opera} designs an open remote attestation TEE platform to prevent the private information from being leaked to TEE platform provider.
However, there is no solution in OPERA design to verify other parties' TEEs are an unexpected platform without leaking other parties' privacy.
S-FaaS \cite{alder2019s} protects the trustworthy and accountable of Function-as-a-Service by setting worker enclaves using Intel SGX.
However, it designs a centralized key distributing enclave to attest these worker enclaves. 
This will collect information on participants who provide functional services.
SRAS does not have these limitations.

Intel SGX provides Enhanced Privacy ID (EPID) \cite{johnson2016intel} scheme and DCAP \cite{dcap} service for enclave remote attestation.  
EPID uses group signatures to prevent privacy leaks, and DCAP provides a more flexible local quote verification scheme.
On top of these solutions, SRAS further explores the issue of multi-party SGX attestation.
Recently, Intel proposes Intel Trust Authority \cite{ITA}, a cloud service that provides attestation service to verify user's workload running in TEE.
However, as disucssed in Sec. \ref{Insights}, it is a centralized service scheme, 
which might not be available due to political policy restrictions in some areas.

\section{Conclusion}\label{conclusion}
We have presented SRAS, a lightweight self-governed remote attestation scheme.
It provides Relying Party with attestation and verification functions to validate the trustworthiness of TEE and compute assets,
achieving a decentralized unified trusted attestation and verification platform for multi-party cloud users.
We have designed an open-source relying party enclave in SRAS, which forms a virtual relying party verifiable network to locally verifying on behalf of the relying parties of the other participants without leaking the sensitive data to others.
We implemented a prototype, analyzed the security of SRAS, and evaluated its performance.







\bibliographystyle{IEEEtranS}
\bibliography{IEEEexample}






\end{document}